\documentclass[aps,pra,showpacs,twocolumn,groupedaddress,lengthcheck]{revtex4}
\usepackage{amsmath}
\usepackage{amssymb}
\usepackage{graphics}
\usepackage[dvips]{graphicx}
\usepackage{graphicx}
\usepackage{color}

\newcommand \beq{\begin{eqnarray}}
\newcommand \eeq{\end{eqnarray}}
\def\simge{\mathrel{%
       \rlap{\raise 0.511ex \hbox{$>$}}{\lower 0.511ex \hbox{$\sim$}}}}
\def\simle{\mathrel{
       \rlap{\raise 0.511ex \hbox{$<$}}{\lower 0.511ex \hbox{$\sim$}}}}

\begin{document}
\title{Landau critical velocity in weakly interacting Bose gases }
\author{Gordon Baym$^{a,b}$ and C.\ J.\ Pethick$^{b,c}$}
\affiliation{\mbox{$^a$Department of Physics, University of Illinois, 1110
  W. Green Street, Urbana, IL 61801} \\
\mbox{$^b$The Niels Bohr International Academy, Blegdamsvej 17, DK-2100 Copenhagen \O,
 Denmark}\\
\mbox{$^c$NORDITA, Royal Institute of Technology and Stockholm University,}\\
\mbox{
Roslagstullsbacken 23, SE-10691 Stockholm, Sweden }\\
}

\date{\today}

\begin{abstract}
    The flow of a uniform Bose gas at speeds greater than the Landau critical velocity, $v_c$, does not necessarily destroy
superfluidity, but rather need only lead to a decrease of the superfluid mass density, $\rho_s$.    Analyzing a weakly interacting Bose gas with a finite range interparticle interaction that leads to a Landau critical velocity at non-zero quasiparticle momentum, we explicitly construct the (non-uniform)
condensate for fluid flow faster than $v_c$ and calculate the accompanying decrease in $\rho_s$.  We briefly comment on the relation of the physics to other problems in superfluids, e.g., solitons, and vortices in Bose-Einstein condensates, and critical currents in superconductors.
\end{abstract}

\maketitle

\section{Introduction}
    
    When a superfluid flows sufficiently rapidly that the excitation spectrum at non-zero momentum in the lab frame becomes gapless, then according to the Landau criterion, superfluidity ceases \cite{landau1941,lev1992}.   However, in many situations, e.g., dilute solutions of $^3$He in superfluid $^4$He, as well as superfluid $^4$He at non-zero temperature,
the excitation spectrum is indeed gapless.  Yet these systems remain good superfluids, only with the superfluid mass density $\rho_s$ less than the total mass density $\rho$.  As a consequence such systems have a non-vanishing normal mass density, 
$\rho_n = \rho-\rho_s$ as well, and thus exhibit two-fluid behavior.  On the other hand, a system with a gap in the excitation spectrum is not guaranteed to be superfluid, as one sees for example in the Mott insulating phase of a Bose gas in an optical lattice \cite{bloch2002}.
In this paper we study in detail the state of superfluidity in a weakly interacting Bose gas flowing at a velocity greater than the Landau critical velocity.    Pitaevskii has laid out the basic physics of this situation in the context of rotons in superfluid $^4$He \cite{lev};  once the rapidly flowing system begins spontaneously to create excitations, interactions between the excitations --when repulsive -- modify the excitation spectrum so that the system remains just critical.   

  We address here flow of the system past walls at speeds $v$ greater than the Landau velocity, $v_c$,  The situation in which an object moves faster than $v_c$ through a superfluid at rest is different:  there the object will generate excitations as it moves and will experience dissipation, as observed in cold atom experiments \cite{ketterle1999}.  The excitations propagate away, and do not affect the energy required for the object to make a further excitations.  
    
\begin{figure}[b]
\begin{center}
\includegraphics[viewport=0 0 423 300,width=10.0cm,height=7cm,clip]{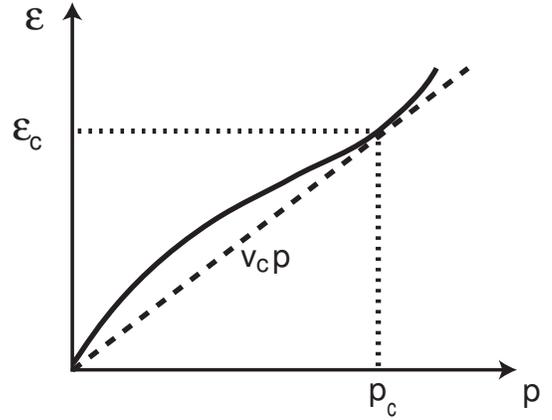}
\caption{The assumed quasiparticle excitation spectrum as a function of momentum.  The slope of the dashed line is the Landau critical velocity, $v_c$.}  
\label{ep}
\end{center}
\end{figure}   

   Explicitly, in a superfluid with an elementary excitation spectrum $\varepsilon(p)$ at momentum $p$ flowing with velocity $v$ (in the positive x direction) with respect to the system walls, the excitation energy in the frame at rest is $\varepsilon(p) + vp_x$.   The Landau criterion states that the superfluid will become energetically unstable when $\varepsilon(p)/p = v$ (for $\vec p\,$ in the negative  x direction, the direction opposite to the flow).  At the minimum $v\equiv v_c$ satisfying this condition, the group velocity of the excitation, $d\varepsilon(p)/dp$, equals its phase velocity $\varepsilon(p)/p$. 
We consider here a weakly interacting Bose gas at zero temperature
initially fully condensed into a state of flow in the x
direction with condensate momentum $q> 0$, and a corresponding order
parameter in the lab frame,
\beq
\psi(x) =\sqrt{n} e^{iqx},
\eeq
with $v=q/m$.
With a zero range interaction, the excitation spectrum curves upward, and the group velocity always exceeds the phase velocity at non-zero momentum. Thus to study the Landau criterion, we assume a finite range interaction $g(|\vec r-\vec r\ '|)$ between particles,  for which the Bogoliubov excitations of the system with momentum $p$ have energy
\beq
\varepsilon(p) = \left[\frac{ng(p) p^2}{m}+\left(\frac{p^2}{2m}\right)^2 \right]^{1/2},
\label{bogq}
 \eeq
 where $n$ is the density and $g(p)$ is the Fourier transform of $g(r)$.
 We take $g(p)$ positive and decreasing with $p$.  Then the phase and group velocities of an excitation are equal when
 \beq
     \frac{nm}{p}\frac{dg(p)}{dp} = \frac{d(gn)}{d(p^2/2m)} = -\frac12.
 \eeq
We assume that this condition is satisfied at the critical momentum $p_c$; thus the flow velocity at  the point at which the Landau criterion is satisfied, is 
\beq
  v _c =  \left[(p_c/2m)^2 + ng_1/m\right]^{1/2} ,
  \label{q}
\eeq
where $g_1 \equiv g(p_c)$, and more generally for integer $\nu$, $g_\nu \equiv g(\nu p_c)$.  The condition that the velocity, $q/m$, exceeds  $v_c$ can be written as
\beq
   q^2 > p_c^2/4 + mng_1.
\eeq  
The excitation spectrum is illustrated in Fig.~1.  Note that it is not necessary for the spectrum to have a (roton-like) minimum in order for there to be a critical velocity at non-zero momentum.   We refer to excitations of momentum near $p_c$ as {\em levons}.

In the next section we lay out the order parameter describing flow faster than the critical velocity, and calculate the corresponding energy of the system.  As we show the order parameter develops an instability at the critical velocity towards formation of additional components at the critical momentum.    In Sec.~III we calculate, for flow velocities above the critical velocity for repulsive inter-excitation interactions and small amplitude modulation of the order parameter, the equilibrium order parameter, the spatially varying density induced by the additional components in the order parameter,  the supercurrent, and the decrease in the superfluid mass density.   We also qualitatively discuss the larger amplitude regime. 
In the final section we relate the present discussion to solitons in Bose-Einstein condensates, to vortices below the critical angular velocity for vortex formation, and to super-critical currents in superconductors, as well as lay out a few possible ways that one could realize non-zero Landau critical velocities in cold atomic gases.

\section{order parameter with flow}

To examine the stability of the system we explore an order parameter in the lab frame,
\beq
\psi(x) = e^{iqx}\left[\sqrt{n_0} +{\cal U} e^{-ip_cx}  -{\cal V}e^{ip_cx}\right],
\label{order}
 \eeq 
corresponding to a reduced number of particles in the initial
 condensate of momentum $q$, and with condensate components of momentum $q\pm p_c$.  The instability occurs,
 as we anticipate, at $p_c >0$, corresponding to an excitation moving in the negative x direction.   We
 work at fixed average density, $\bar n$.  Such an order parameter, for small
 ${\cal U}$ and ${\cal V}$ describes the usual Bogoliubov excitations of the condensate.
 While ${\cal U}$ and ${\cal V}$ can be generally complex, the most energetically favorable situation
corresponds to ${\cal U}$ and ${\cal V}$ real, as will be apparent once we calculate the energy corresponding to the state (\ref{order}).  Without loss of
generality, we take ${\cal U}\ge0$, and ${\cal U}^2\ge {\cal V}^2$.

The number density is
\beq
n(x) &=& |\psi(x)|^2 =n_0 + {\cal U}^2 +{\cal V}^2\nonumber\\ &&+ 2\sqrt{n_0}({\cal U}-{\cal V})\cos p_cx - 2{\cal U}{\cal V}\cos 2p_cx,
\label{nx}
\eeq
and the spatial average of the density is
\beq
\bar n = n_0 + {\cal U}^2+{\cal V}^2.
\eeq
Thus ${\cal U}^2+{\cal V}^2$ is bounded above by $\bar n$.  In the following we drop the bar over the $n$ to simplify the notation.
Similarly, the average kinetic energy density is
\beq
E_{ke} &=& \frac{1}{2m}\left(n_0q^2 + {\cal U}^2(q-p_c)^2
 +{\cal V}^2(q+p_c)^2\right)\nonumber\\
 &=& \frac{1}{2m}\left(n q^2 + ({\cal U}^2+{\cal V}^2) p_c^2 - 2({\cal U}^2-{\cal V}^2)qp_c \right).
\eeq
The average interaction energy density is 
\beq
E_{int} = \frac1{2\Omega} \int d^3r d^3r' g(|\vec r - \vec r\ '|)
|\psi(x)|^2|\psi(x')|^2
\nonumber \\
= \frac12 g_0 n^2 +  g_1n_0 ({\cal U}-{\cal V})^2 + {\cal U}^2{\cal V}^2 g_2,
\eeq
where $\Omega$ is the system volume \footnote{One cannot gain energy by allowing ${\cal U}$ and ${\cal V}$ to be complex: ${\cal U}\to {\cal U}e^{i\theta_u},  {\cal V}\to {\cal V}e^{i\theta_v}$.  Since the phase difference of ${\cal U}$ and ${\cal V}$ can be absorbed in the choice of origin of $x$, we need only consider the more general order parameter
$
\psi(x) = e^{iqx}\left[\sqrt{n_0} + e^{i\theta}\left({\cal U} e^{-ip_cx} - {\cal V}e^{ip_cx}\right)\right]
$.
The net effect on the energy is to replace the $({\cal U}-{\cal V})^2$ in Eq.~(\ref{eprime}) by ${\cal U}^2+{\cal V}^2-2{\cal U}{\cal V}\cos\theta$.   This final term
is most negative however for $\theta = 0$.   Thus allowing phase variations of ${\cal U}$ and ${\cal V}$ cannot decrease the energy.}. 
We note that only the $({\cal U}-{\cal V})^2$ term in the energy distinguishes the relative sign of ${\cal U}$ and ${\cal V}$, and this term, for given ${\cal U}$ and $|{\cal V}|$ favors ${\cal V}$ having the same sign as ${\cal U}$.  In addition the mass current,
\beq
mj &=& qn_0 + (q-p_c){\cal U}^2 + (q+p_c) {\cal V}^2  \nonumber\\ 
    &=& qn - ({\cal U}^2-{\cal V}^2)p_c,\label{mass}
\eeq
is reduced from $qn$ for $p_c>0$.

   To analyze the stability of the system we write
\beq
{\cal U} =  \zeta \cosh(\eta/2),\quad\quad {\cal V} =  \zeta\sinh(\eta/2),
\eeq
where $\zeta \ge 0$, and $0\le \eta\le \infty$.  Then the average number density is
\beq
  n = n_0 + \zeta^2 \cosh \eta,
\eeq
which implies the bound, $\zeta^2\cosh\eta\le n$.   In terms of $\zeta$ and $\eta$ the average energy density is
\beq
E &=& E_{ke}+E_{int}= \frac12 g_0n^2 + \frac{q^2}{2m}n +E',
 \label{e}
\eeq
where
\beq
  E' &=&\zeta^2\left(\frac{p_c^2}{2m}\cosh\eta
  -\frac{p_cq}{m} + g_1 n e^{-\eta}\right) \nonumber\\
  &&+\zeta^4 \left(\frac14 g_2 \sinh^2 \eta - g_1 e^{-\eta} \cosh\eta \right).
  \label{eprime}
\eeq
We suppress the two constant terms in Eq.~(\ref{e}) which play no role, and work with the energy density $E'$.

Let us first ignore the $\zeta^4$ terms; minimizing the first line in (\ref{eprime}) with respect to $\eta$ at fixed $\zeta$,  we find \footnote{The quantity  $e^{-\eta/2}$ corresponds in Pitaevskii's calculation \cite{lev} to the modulus of the matrix element of the density operator between the uniform ground state and the state with one critical excitation.}:
\beq
   e^{-\eta} = \frac{({\cal U}-{\cal V})^2}{\zeta^2} = \frac{p_c^2}{2m\varepsilon_c},
\label{phi}
\eeq
where
\beq
  \varepsilon_c =  \left[\frac{p_c^2}{2m}\left(\frac{p_c^2}{2m}+2g_1n\right)\right]^{1/2};
\eeq
is the usual Bogoliubov expression for an excitation of momentum $p_c$.  For $g_1 > 0$, $\eta$ is positive.

With this solution,
\beq
  E' = \zeta^2\left(\varepsilon_c - \frac{qp_c}{m}\right).
\eeq
We see explicitly that for $q < mv_c$, the Landau critical momentum, the original superfluid flow is stable against small excitation as expected, but for $q>mv_c$ the system is unstable against  developing a non-uniform condensate with $\zeta \ne 0$.  The instability sets in, as noted, for $p_c>0$, corresponding to an excitation moving in the negative x direction.
For $\zeta\ll1$,  the system energy $E'$ is positive for all $q<mv_c$.   For
$q\,\ge\, mv_c$, however, the energy as a function of $\zeta$ decreases steadily
from zero indicating a transition to a state of nonzero $\zeta$.

\section{above the critical velocity}

   We now examine the state of the system for flow velocity slightly above the critical velocity $v_c$.  While $\partial E/\partial \zeta^2$ is initially negative, the terms of order $\zeta^4$ can, depending on parameters, locally stabilize the system at small $\zeta^2$.  To order  $\zeta^4$, we may take $\eta$ to be given by Eq.~(\ref{phi}), since the $\zeta^2$ corrections to $\eta$ do not contribute to $E'$ at the minimum with respect to $\eta$.   The terms to order  $\zeta^4$ in $E'$ become
   \beq
E' &=& \zeta^2\left(\varepsilon_c - \frac{qp_c}{m}\right)+\frac{\gamma}{2}  \zeta^4
\label{ez4}
\eeq  
where $\gamma$, given by
\beq
\gamma = \frac{g_1}{2\varepsilon_c^2}
\left[g_1g_2 n^2 - \frac{p_c^2}{m}\left(\frac{p_c^2}{m}+2g_1n\right)\right],
\eeq
corresponds,
in terms of Pitaevskii's description of He-II above the Landau critical velocity \cite{lev},
to the roton-roton interaction.   Note that for small $g_2$, $\gamma$ is negative, while for 
\beq
   \frac{g_2}{g_1} >    
     \frac{p_c^2}{m(g_1n)^2}\left(\frac{p_c^2}{m}+2g_1n\right),
\eeq 
$\gamma>0$,  and the system has a local minimum at
\beq
 \zeta^2 = (v-v_c)\frac{p_c}{\gamma}.
 \label{z2} 
\eeq
This minimum corresponds to a nonzero condensate amplitude at momentum $p_c$.  The excitations near $p_c$ form a levon condensate, or in the context of $^4$He, a ``Bose condensation'' of rotons \cite{iordanskii}.  

   The variation of the energy with $\zeta^2$ about this minimum, which vanishes by construction, defines the renormalized energy of an excitation of momentum $p_c$ in the lab frame,
\beq
   \frac{\partial E'}{\partial\zeta^2} = \varepsilon_c + \zeta^2\gamma -
\frac{qp_c}{m} = 0.
\label{dedz2}
 \eeq
This structure agrees
with Pitaevskii's picture that the roton spectrum in the frame moving with the superfluid is shifted by the presence of critical excitations, 
$\varepsilon_c \to \varepsilon_c + \zeta^2 \gamma \equiv \varepsilon_{c,R}$, and
the renormalized excitation energy in the lab frame, $\varepsilon_{c,R} - qp_c/m$, vanishes.  

  Furthermore the density has a ``layered structure," Eq.~(\ref{nx}), given to first order in $\zeta$ by
\beq
n(x) = n +2\sqrt n_0 \zeta\cosh\eta\cos(p_cx).
\eeq
This structure,  first predicted by Pitaevskii, has been seen in a density-functional simulation of flow in $^4$He, with amplitude $\sim (v-v_c)^{1/2}$, corresponding to
Eq.~(\ref{z2}) \cite{ancilotto}. 
The total mass current in this state with nonzero $\zeta$ is, from Eq.~(\ref{mass}),
\beq
mj = \partial E/\partial q = qn -p_c\zeta^2.
\eeq
Thus slightly above the Landau critical velocity, the superfluid mass density is,
\beq
  \rho_s \equiv \frac{mj}{q} \simeq mn - \frac{p_c^2}{\gamma}\left(\frac{v}{v_c} -1   \right),
 \eeq
Superfluidity does not cease at the Landau critical velocity.  Rather, as we
see in this model calculation, as the Landau critical velocity is
exceeded the superfluid mass density begins to fall below the total mass
density,  here linearly with the flow velocity.  The reduction in $\rho_s$ can be understood as a consequence of the lack of translational invariance of the system above $v_c$ \cite{ptr}.  Furthermore, the normal mass density is
\beq
  \rho_n =  \frac{p_c}{v_c}\zeta^2   = \frac{p_c^2}{\gamma}\left(\frac{v}{v_c} -1   \right).
\eeq
Quite generally, the superfluid mass density need not fall abruptly to zero at the critical velocity.

 The superfluid above the critical velocity has a richer mode
structure than below.  Ordinary zero sound -- the conventional
Bogoliubov excitations -- should not be
modified significantly for flows only slightly above $v_c$.  On the
other hand, the presence of the normal component (the levon condensate)
will lead to a additional sound-like mode, due to variations of the levon
density and mean momentum.   Noting that the calculations leading to Eq.
(23) do not depend on $q_c$ being the critical momentum, we infer that the
energy of an excitation of momentum $p$ is given by
\beq
 \omega_p^0 = \varepsilon_p+\zeta^2\gamma - qp/m.
\label{wp}
\eeq
More generally, levons with momentum $p$ are mixed with those with
momentum $2p_c-p$ because a pair of such excitations can be annihilated,
thereby producing two levons with momentum
$p_c$.  This process may also be considered as an Umklapp process
resulting from the periodic lattice structure. As Pitaevskii showed
\cite{lev} for excitations of the ``roton condensate" in superfluid
$^4$He, the excitation energies of the modes with momentum near the
levon momentum for arbitrary direction of $\vec p$ are given more generally by
\beq
   \omega_{\vec p} = \left[2\zeta^2\gamma (\varepsilon_p -\vec q\cdot \vec p\,/m)  +
(\varepsilon_p -\vec q\cdot \vec p\,/m)^2)\right]^{1/2},
\eeq
which reduces to the above result (\ref{wp}) for small positive $v -
v_c$, and $\vec p$ along $\vec q$.  The levon energy in the lab frame is 
\beq
  \omega_{\vec p}^2\sim (v-v_c)p_c\left(\frac{Q_\parallel^2}{\mu_\parallel} +  \frac{Q_\perp^2}{\mu_\perp} \right)
 \eeq 
for small $\vec p - \vec p_c$, where $Q_\parallel = p_x - p_c$, $Q_\perp = (p_y,p_z)$ and the $\mu$'s are effective masses.  The structure of the levons is simply that of phonons of the layered condensate. 
Because the  levon energy depends on the local particle density
and superfluid velocity, these modes will be hybridized with ordinary
Bogoliubov zero sound excitations, thereby inducing an attractive
interaction between levons.  More detailed calculations of mode
frequencies can be made using the methods employed for a Bose-Einstein
condensate in an optical lattice  \cite{WuNiu,machholm}, the only
difference here being that the spatial modulation of the density is due
to the flow, not an external potential.

Let us look at the energy landscape more globally.  For $v$ slightly above $v_c$ and $\gamma>0$ the system has a stable local minimum at small nonzero ${\cal U}$ and ${\cal V}$.   When $\gamma <0$, there is no local minimum at small $\zeta^2$; were the term of order $\zeta^6$ in the free energy positive, the system would undergo a discontinuous transition, leading to a discontinuity in $\rho_s$.
However, to describe the most favorable state of the system for $\gamma<0$ requires that one goes beyond using the simple trial wave function, Eq.~(\ref{order}).  Even for 
 $p_c/2<q < mv_c$, the energy $E'$ initially rises with increasing $\zeta$, indicating stability of the flow under small perturbations, but falls through zero at a nonzero $\zeta$ determined from\footnote{The state of least energy for $q>p_c/2$ is always at $\eta=0$ and $n_0=0$, corresponding to a condensate of momentum $q-p_c$, but this state is not germane to the present discussion.   
This result is independent of the interactions; inclusion of the $g_2$ term in the energy cannot provide absolute stabilization of the system at nonzero $\zeta$.}  Eq.~(\ref{eprime}).  On the other hand, for $q<p_c/2$ the energy $E'$ is positive for all nonzero $\zeta$, indicating absolute stability of the flow.

\section{discussion}

The wave function (\ref{order})  for nonzero ${\cal U}$ and ${\cal V}$, the lowest energy state at velocities greater than $v_c$, may be regarded as a periodic array of solitons.  Such arrays were first considered for the Gross--Pitaevskii equation  by Tsuzuki \cite{tsuzuki} and they have been studied for the case of a periodic background potential by Machholm et al. \cite{machholm}, who showed that they correspond to states on the upper branch of the ``swallow tail" in the spectrum.   In these works the solitons were found to have an energy {\it greater} than that of the uniform state moving at constant velocity.  The new feature of the situation considered in the present paper is that for velocities greater than $v_c$, the periodic array of solitons has an energy {\it less} than that of the uniform state.  This is a consequence of the nonzero range of the potential, which here makes density fluctuations with wavevectors close to $p_c$ less energetically costly than long-wavelength density fluctuations having the same amplitude. 

An approach similar to the present can be used to describe instabilities of vortices in Bose condensates.  Reference~\cite{soheil} considered a single vortex in a trap as a function of the rotation frequency, $\Omega$.  For example, for $\Omega$ less than the critical frequency $\Omega_c$ at which the system can first support a vortex at the center, the vortex is not stable against fluctuations, but has a negative energy anomalous mode \cite{anom}.  The situation is analogous to that in the present paper above the critical velocity.  The condensate wave function (cf.~Eq.~(\ref{order}))
$ \psi \sim \sqrt{n_1} e^{i\phi} + u + v e^{2i\phi}$,
where $\phi$ is the azimuthal angle, then has lower energy, and describes two off-center vortices in the cloud asymmetric about the origin.
 
It is interesting to compare the present situation for bosons with 
the analogous phenomenon in superconductors.    As shown
by Rogers \cite{rogers} (quoted in \cite{bardeen}), when the superfluid velocity exceeds the Landau critical velocity, $v_c = \Delta/p_F$, where $\Delta$ is the gap and $p_F$ the Fermi momentum, the system spontaneously generates quasiparticles.  At zero temperature each quasiparticle state of negative energy becomes occupied; the Pauli principle here plays the role of stopping the process from running away.   Although the current continues to increase initially with increasing flow velocity above $v_c$, it quickly goes though a maximum and then goes to zero at velocity $(e/2)v_c$, where $e = 2.718\dots$ is Napier's constant.  In contrast to the Bose case, the excitations are fermions, so they cannot all be in the same momentum state.  As a consequence, there is no density modulation for flow velocities greater than $v_c$.

In this article we have only studied states in which the amplitude of the spatial non-uniformity is small: in the future it would be interesting to extend this work to  larger amplitude disturbances, including the possibility that the order parameter could develop nodes, and even vortex lines could be formed.   Numerical simulations in this non-linear regime would be very helpful.  A further problem is to determine the state of the system at finite temperature, including fluctuations of the condensate structure from the levons.

Finally, let us ask whether one can realize the present model experimentally.  A first way could be to take advantage of enhancement of the effective range near a Fesh\-bach resonance.  Essentially, for scattering of two particles at relative momentum $p$ with total momentum 0, the contribution of a resonance to the effective interaction, $g(p)$, is
\beq
      g(p)_{res} \simeq \frac{|M|^2}{\Delta E+ p^2/4m}.
\eeq
where $\Delta E$ is the energy difference of the particles of zero relative momentum in the entrance and intermediate channels, and $|M|^2$ is the square of the matrix element between the two channels.   Thus 
\beq
   n\frac{d g(p)}{d(p^2/2m)} \simeq -\frac{n}{2}\frac{|M|^2}{(\Delta E+ p^2/4m)^2},
\eeq  
which can reach the needed value $ -1/2$ sufficiently close to the Feshbach resonance at small $p$.  
A second possibility could be to use the downward bending of the  quasiparticle energy $\varepsilon(p)$ with increasing $p$, arising from short range correlations.  To the extent that the f-sum rule is exhausted by a single quasiparticle, the quasiparticle energy is given by $p^2/2mS(p)$, where $S(p)$ is the static structure factor \cite{feynman}; quite generally short range correlations tend to enhance $S(p)$ at shorter wavelengths -- in $^4$He leading to the roton dip, as Feynman first argued --  and could produce the needed softening of the spectrum.
A third possibility could be to use atoms with dipole-dipole interactions in quasi-one dimensional systems.   We leave the elucidation of these methods to a future study.

As we have shown in the calculations in this paper for weakly interacting condensed bosons, a uniform flow at velocity exceeding the Landau critical velocity does not necessarily destroy the superfluidity.   For positive interactions among the critical excitations, the system becomes stabilized with a reduced  superfluid mass density.   A small amplitude analysis for negative interactions among the quasiparticles is however inadequate, and  elucidating the possible states to which the system might go requires further work.

\section*{Acknowledgements}

Author GB is grateful for the hospitality of the Aspen Center for Physics, where part of
this work was carried out.  We thank Tony Leggett and Soheil Baharian for critical comments.
This research was supported in part by NSF Grants PHY07-01611 and PHY09-69790.

\end{document}